# Contention-based Grant-free Transmission with Extremely Sparse Orthogonal Pilot Scheme

Zhifeng Yuan[1,2], Zhigang Li[1,2], Weimin Li[1,2], Yihua Ma[1,2]
[1]ZTE Corporation, Shenzhen, China
[2]State Key Laboratory of Mobile Network and Mobile Multimedia Technology, Shenzhen, China
Email: {yuan.zhifeng, li.zhigang4, li.weimin6, yihua.ma}@zte.com.cn

*Abstract*—Due to the limited number of traditional orthogonal pilots, pilot collision will severely degrade the performance of contention-based grant-free transmission. To alleviate the pilot collision and exploit the spatial degree of freedom as much as possible, an extremely sparse orthogonal pilot scheme is proposed for uplink grant-free transmission. The proposed sparse pilot is used to perform active user detection and estimate the spatial channel. Then, inter-user interference suppression is performed by spatially combining the received data symbols using the estimated spatial channel. After that, the estimation and compensation of wireless channel and time/frequency offset are performed utilizing the geometric characteristics of combined data symbols. The task of pilot is much lightened, so that the extremely sparse orthogonal pilot can occupy minimized resources, and the number of orthogonal pilots can be increased significantly, which greatly reduces the probability of pilot collision. The numerical results show that the proposed extremely sparse orthogonal pilot scheme significantly improves the performance in high-overloading grant-free scenario.

*Keywords—grant-free transmission, extremely sparse orthogonal pilot scheme, pilot collision*

## I. Introduction

Massive machine-type communication (mMTC), as one of the important application scenarios in future wireless networks, has attracted extensive attention from academia and industry [1]. In mMTC, the main challenge is to support massive user equipments (UEs), e.g. the IoT devices, in uplink transmission with sporadic small data payload, low cost, and low complexity. To address this challenge, grant-free non-orthogonal multiple access (NOMA) has been considered as a promising technique [2], [3]. Grant-free transmission allows UEs to transmit data payloads autonomously without the need to send scheduling request and wait for dynamic scheduling. NOMA can increase the user loading by allowing UEs to share the same time and frequency resource via power/code domain multiplexing. The combination of grant-free and NOMA can significantly reduce signaling overhead, terminal complexity, and power consumption while achieving a higher user loading compared with the grant-based orthogonal transmission, which is very suitable for mMTC scenario.

However, as there is no scheduling of the transmission resource by the base station (BS), the grant-free transmission is contention-based. Take the crucial pilot signal as an example, active UEs in grant-free transmission autonomously select pilot sequences (e.g. demodulation reference signal (DMRS) sequences or preambles) from the predefined pilot sequence set. Inevitably, multiple active UEs may select the same pilot sequence, which is called 'pilot collision' [4]. For a given pilot set, the probability of pilot collision increases rapidly with the number of active UEs. Pilot collision will lead to miss detection and inaccurate channel estimation of collided UEs, which severely degrades both the suppression of inter-user interference (IUI) and the compensation of channel distortion experienced by the data symbols. Moreover, due to the extremely simple transmit procedure of contention-based grant-free (CBGF) transmission, the received signals could experience large time offset (TO) and frequency offset (FO): 1) Due to the lack of uplink timing alignment/timing advance (TA) procedure, the transmitted signals of active UEs may arrive at the BS with different time delays, with each UE's delay being determined by its distance to the BS, thus the received signals from the UEs near the edge of the cell would experience large TOs [5]. 2) Due to the lack of tight frequency synchronization, the oscillator misalignment and Doppler effect could cause a large FO [6]. Large TO/FO will further increase the symbol distortion on the basis of distortion induced by wireless multipath channels, which makes channel estimation and symbol demodulation more difficult. As a result, it's very challenging for the multi-user detection (MUD) of CBGF transmission as it will confront not only heavy IUI and severe distortion on the received symbols, but also uncontrollable pilot collision. To achieve a better MUD performance for the CBGF transmission, different pilot designs and transceivers have been proposed. Orthogonal pilot, following the existing DMRS design in 5G new radio (NR), is employed [7]. In the following, we call the grant-free transmission using existing orthogonal pilot design as the traditional orthogonal pilot (TOP) scheme. It should be noted that the orthogonal pilot used in TOP scheme needs to have enough density in both time and frequency domain to estimate the wireless channel and TO/FO, thus only limited number of orthogonal pilots can be provided under certain pilot overhead, resulting in severe pilot collision in high-loading scenario. It was confirmed that the number of orthogonal pilots dominates the success probability [8]. Hence, TOP scheme in general is not optimal for CBGF transmission due to the limited number of pilots and severe pilot collision. Non-orthogonal pilot based grant-free transmission has also been widely studied as it can provide a much larger number of pilots than TOP and thus alleviate the pilot collision. The BS can pre-configure one unique non-orthogonal pilot sequence for each UE [9], or each active UE can autonomously select one pilot sequence from the predefined non-orthogonal pilot set [10]. Typically, compressive sensing based active user detection and channel estimation are utilized at the receiver, as only a small proportion of potential UEs are active and transmitting data each time [11]. Although non-orthogonal pilot based scheme can alleviate pilot collision compared with TOP, the mutual interference of non-orthogonal pilots will degrade the

This work was supported by the National Key Research and Development Program of China under Grant 2020YFB1807202.

accuracy of channel estimation. What's more, the estimations of TO/FO using non-orthogonal pilots haven't been well solved yet.

In [12], [13], data-only grant-free scheme without pilot signal was proposed, which can eliminate the pilot collision and overhead. Different from pilot-based scheme, data-only scheme first performs IUI suppression by fully blind spatial and/or code domain combination. Then the estimations of the symbol distortions induced by the wireless channel and TO/FO are performed by utilizing the characteristics of the combined data symbols. However, data-only scheme will face the problem of phase ambiguity and it is not easy to design high-performance blind spatial combination vectors when the BS is equipped with large-scale antennas with low spatial correlation.

In this paper, taking advantage of both pilot-assisted and data-only schemes, a novel extremely sparse orthogonal pilot (ESOP) scheme is proposed to handle the aforementioned problems. First, active user detection and estimation of the active UEs' spatial channels are performed using the proposed ESOP in a simple way of correlation detection. Next, the spatial domain suppression of IUI can be carried out by spatial combination using the estimated spatial channels. Then, the estimations of symbol distortions induced by the wireless channel and TO/FO are carried out by utilizing the geometric characteristics of combined data symbols [12], [13]. It can be seen that the responsibility of ESOP is much lightened than TOP as some important estimations are done by the data symbols, thus, the pilot sequences of ESOP are no longer required to be spread out in both time and frequency domain and can occupy minimized resources. Under the same pilot overhead, the number of pilot sequences in the ESOP can be maximized, therefore, the pilot collision can be greatly alleviated compared with TOP scheme. Based on [7], ESOP-based independent multi-pilot scheme is proposed, and the probability of pilot collision can be decreased further. The link-level simulation results show the performance of the proposed ESOP scheme is greatly improved compared with TOP scheme. What's more, the proposed ESOP scheme can work well when larger TO and FO exist.

The rest of this paper is organized as follows. In Section II, the system model and existing works are introduced. In Section III, the proposed ESOP scheme, including the pilot design, the analysis of pilot collision probability, and the receiver flow, is introduced. Section IV provides the performance evaluation results together with the comparisons with TOP scheme. The conclusion is presented in Section V.

*Notations*: Boldface lower and upper case symbols represent vectors and matrices, respectively. $\mathbf{p}^t$ and $\mathbf{p}^*$ denote the transpose and complex conjugate-transpose of $\mathbf{p}$, respectively.

## II. SYSTEM MODEL AND EXISTING WORK

*A. System Model*

We consider an uplink grant-free transmission scenario consisting of one BS equipped with $M$ receive antennas and massive potential single-antenna UEs. The predefined pilot set $\mathbf{P}$ has $N$ orthogonal pilot sequences of length-$N$, and can be denoted as $\mathbf{P} = [\ \mathbf{p}_1^t, \mathbf{p}_2^t, \ldots, \mathbf{p}_N^t\ ]^t$, where $\mathbf{p}_n \in \mathbb{C}^{1 \times N}$, $n = 1, \ldots, N$. When data traffic arrives, active UE autonomously selects one pilot sequence from the predefined orthogonal pilot set and transmits it with data payload to the BS. Assume $K$ active UEs, the received superimposed pilot symbols can be expressed as

$$\mathbf{y}_P = \sum_{k=1}^{K} \mathbf{h}_k \mathbf{p}_k + \mathbf{n}_P \quad (1)$$

where, $\mathbf{h}_k \in \mathbb{C}^{M \times 1}$, $\mathbf{p}_k, \mathbf{y}_P \in \mathbb{C}^{M \times N}$, $\mathbf{n}_P \in \mathbb{C}^{M \times N} \sim \mathcal{CN}(0, \sigma^2)$ denote the channel vector and pilot sequence of the $k$-th UE, the received superimposed pilot symbols, and the complex Gaussian noise, respectively.

Similarly, the received superimposed data symbols can be expressed as

$$\mathbf{y}_d = \sum_{k=1}^{K} \mathbf{h}_k \mathbf{s}_k + \mathbf{n}_d \quad (2)$$

where, $\mathbf{s}_k \in \mathbb{C}^{1 \times L}$, $\mathbf{y}_d \in \mathbb{C}^{M \times L}$, $\mathbf{n}_d \in \mathbb{C}^{M \times L} \sim \mathcal{CN}(0, \sigma^2)$ are respectively the transmitted data symbols of $k$-th UE, the received superimposed data symbols, and the complex Gaussian noise. $L$ is the length of transmitting data symbols.

*B. Existing Work*

In this subsection, TOP scheme using existing orthogonal pilot design is introduced. The use of orthogonal pilot allows the BS to produce independent channel estimations for the active UEs, which is needed for demodulation and IUI suppression. Although non-orthogonal pilot based scheme can generally alleviate pilot collision compared with TOP, the non-orthogonality of pilot will degrade the accuracy of channel estimation. Moreover, the estimation and correction of TO/FO using non-orthogonal pilot haven't been well solved yet, therefore, non-orthogonal pilot based scheme is not within the comparison of this paper.

In [7], DMRS configuration type 2 of 5G NR is adopted. The DMRS in 5G NR is generated using pseudo-random sequence. Taking DMRS configuration type 2 occupying two OFDM symbols, as shown in Fig. 1(a), as an example, three code division multiplexing (CDM) groups with each containing four orthogonal reference signals (or orthogonal antenna ports) can be created over the entire transmission bandwidth, thus 12 orthogonal reference signals can be provided at the overhead of 1/7 transmission resources. Enhanced DMRS design has also been considered to increase the number of orthogonal reference signals in existing protocol. In enhanced DMRS design, by reducing the DMRS density in frequency domain, six CDM groups can be achieved, providing up to 24 orthogonal reference signals. Based on the enhanced DMRS design, independent multi-pilot scheme was proposed [7]. Theoretical analysis showed that the probability of pilot collision of independent multi-pilot scheme can be further reduced. Although improved performance can be achieved according to the simulation results, the performance of independent multi-pilot scheme is still restricted by the small pilot set based on the TOP.

The dense pilot design similar to Fig. 1(a) can estimate the wireless multipath channels and TOs of active UEs, but it is not usually possible to estimate their FOs. To estimate the FO, one method is to place a copy of the same pilot at a certain interval in the time domain, which undoubtedly increases pilot overhead. Given a certain pilot overhead, the number of available orthogonal pilots is very limited if TOP scheme is employed, and it is difficult to support a high-loading CBGF transmission. Increasing the pilot overhead can increase the number of orthogonal pilots. However, the spectrum efficiency is reduced as the resource for data transmission is less.

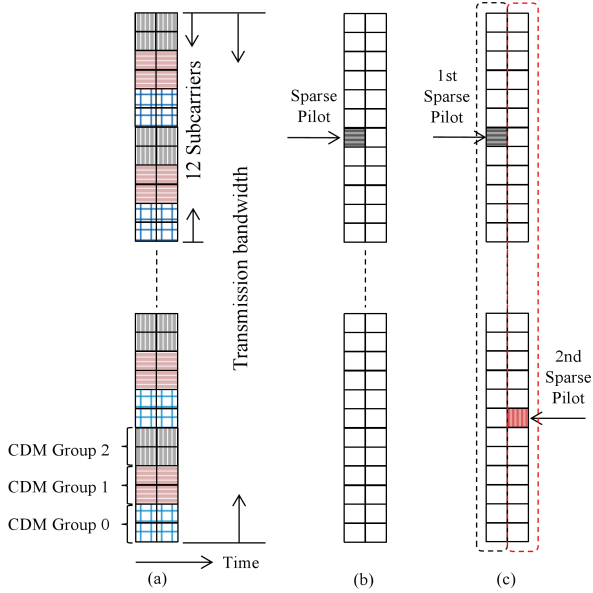

Fig. 1. The structure of (a) existing orthogonal pilot used in TOP scheme; (b) the proposed ESOP scheme; (c) the proposed ESOP scheme configured with 2 independent pilots.

III. THE EXTREMELY SPARSE ORTHOGONAL PILOT SCHEME

*A. Design of ESOP*

The proposed ESOP scheme greatly lightens the task of the pilot, such that each pilot can occupy minimal resources, or be most sparse, resulting in a maximum number of orthogonal pilots. One typical configuration is shown in Fig. 1(b), the proposed ESOP has only one non-zero symbol on one resource element (RE, one subcarrier in frequency domain, and one OFDM symbol in time domain), and there is no signal in the remaining pilot resources (or the value is 0). Assume that the transmission resource is 6 physical resource blocks (PRBs, 12 subcarriers in frequency domain, and 14 OFDM symbols in time domain for normal CP), and each PRB has 12*14 REs. The pilot overhead is the same as in Fig.1(a), i.e., a total number of 144 REs are reserved for pilot transmission under 1/7 overhead, therefore, 144 orthogonal pilots are available for the proposed ESOP scheme, which is at least 6 times that of TOP scheme. Different sparse pilots can be distinguished simply by the positions of non-zero element. The structure of sparse pilot can be configured flexibly, other configurations, e.g. the CDM group in Fig. 1(a), can also be used.

*B. Probability of Pilot Collision*

In grant-free transmission, active UE transmits the pilot sequence autonomously selected from the predefined pilot set. To evaluate the probability of pilot collision, we assume that if the pilot sequence selected by one UE does not collide with others', this UE can be successfully decoded. Otherwise, if more than one UE selects the same pilot sequence, pilot collision occurs and all these collided UEs would not be decoded correctly. Here, the probability of pilot collision is calculated as the ratio of the number of UEs not decoded correctly to the number of total active UEs.

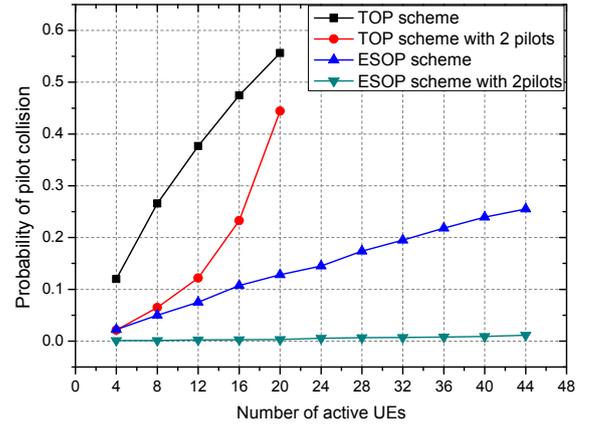

Fig. 2 The probability of pilot collision

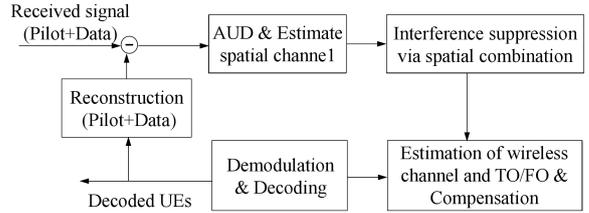

Fig. 3.The receiver of proposed ESOP scheme

The probability of pilot collision is obtained via simulation for $10^3$ times. 6 PRBs and 1/7 pilot overhead are assumed. The probability of pilot collision of both TOP scheme and the proposed ESOP scheme are compared. In addition, independent multi-pilot scheme in [7] is also considered. For TOP scheme and the proposed ESOP scheme, 24 and 144 orthogonal pilots can be provided, respectively. For independent multi-pilot scheme, assuming that both TOP and ESOP scheme are configured with two independent pilots, the two independent pilots can be time/frequency/code division multiplexed. Fig. 1(c) shows time division multiplexing of two independent pilots, and the pilot transmission resource for each pilot is halved. Therefore, for both TOP and ESOP scheme configured with 2 pilots, the size of pilot set is halved to 12 and 72, respectively. As shown in Fig. 2, for TOP scheme, the probability of pilot collision rises rapidly as the number of active UEs increases. In contrast, the probability of pilot collision of the proposed ESOP scheme is much lower than that of TOP scheme, thanks to the greatly increased number of pilots. Moreover, both schemes configured with two independent pilots can further alleviate the pilot collision.

*C. Receiver Flow*

A receiver based on iterative multi-user detection with interference cancellation (IC) can be adopted, as shown in Fig. 3. In the proposed ESOP scheme, the task of pilot is to perform active user detection and estimate the spatial channels of the active UEs, which are used to generate the spatial combination vectors. As the pilots are orthogonal, by simply correlating the received pilot symbols $\mathbf{y}_P$ with all the pilot sequences in the pilot set $\mathbf{P}$, the estimated spatial channels can be expressed as

$$\hat{\mathbf{h}}_n = \mathbf{h}_n + \frac{\mathbf{n}_P \mathbf{p}_n^*}{\mathbf{p}_n \mathbf{p}_n^*}, \quad n = 1, \ldots, N \qquad (3)$$

By setting appropriate detection threshold T, active UEs can be determined. If $|\hat{\mathbf{h}}_n|^2 > T$, then the receiver can determine that one active UE selects the $n$-th pilot sequence $\mathbf{p}_n$. It should be noted that, if only one UE selects the $n$-th pilot sequence $\mathbf{p}_n$, then this UE does not collide with others, and $\hat{\mathbf{h}}_n$ is an accurate estimation of its spatial channel; if two or more UEs select the $n$-th pilot sequence $\mathbf{p}_n$, then the receiver will treat these collided UEs as one, because only one pilot sequence $\mathbf{p}_n$ is identified, and this inaccurate spatial channel estimation $\hat{\mathbf{h}}_n$ is the sum of these collided UEs' spatial channels.

Assume that $Q$ active UEs are detected with a given detection threshold, and the estimated spatial channel matrix is $\hat{\mathbf{H}} = [\hat{\mathbf{h}}_1, \hat{\mathbf{h}}_2, \ldots, \hat{\mathbf{h}}_Q]$. Based on $\hat{\mathbf{H}}$, IUI suppression can be done via appropriate spatial combination. For the $q$-th identified active UE, the spatially combined data symbols can be expressed as

$$\hat{\mathbf{s}}_q = \mathbf{w}_q \mathbf{y}_d, \quad q = 1, 2, \ldots, Q \quad (4)$$

where, $\mathbf{w}_q$ is the spatial combination vector of the $q$-th active UE. If minimum mean square error (MMSE) based spatial combination is used, then

$$\mathbf{w}_q = \mathbf{h}_q^*(\hat{\mathbf{H}}\hat{\mathbf{H}}^* + \sigma^2 \mathbf{I})^{-1} \quad (5)$$

where, $\mathbf{I}$ is the identity matrix. Taking into account pilot collisions, Eq. (5) is no longer a standard MMSE. The impact of pilot collisions can be found in [14], which shows that the performances of both collided and uncollided users degrade. Therefore, when pilot collisions are frequent, the MMSE performance cannot be ensured. To achieve better performance in high loading scenario, an approximation approach is introduced [13]

$$\mathbf{R}_y := E(\mathbf{y}_d \mathbf{y}_d^*) \approx \hat{\mathbf{H}}\hat{\mathbf{H}}^* + \sigma^2 \mathbf{I} \quad (6)$$

where, $\mathbf{R}_y$ is the auto-correlation matrix of the received superimposed data symbols $\mathbf{y}_d$. Hence, Eq. (5) can be approximated as

$$\mathbf{w}_q \approx \mathbf{h}_q^*(\mathbf{R}_y)^{-1} \quad (7)$$

Eq.(5) and Eq. (7) have different working principles, thus perform differently in diverse cases. The better one will be selected by the proposed scheme to support more users.

For TOP scheme, the wireless channel and TO/FO can be well estimated because the pilot is spread over the entire transmission resource. The main factor that restricts the CBGF performance is the severe pilot collision caused by the limited number of pilots. In contrast, the ESOP scheme has much lower pilot collision probability in CBGF transmission, but it also has a penalty: the sparse pilot can be used to estimate the spatial channel of the subcarrier occupied by it, however, the spatial channels of the subcarriers far away from the pilot in general have some deviation from this estimation. Such that the spatial combination using these estimations is usually not optimal for the data symbols in subcarriers far away from the ESOP, leading to a degraded combination performance. In general, there is a trade-off between the estimation accuracy per pilot and the pilot collision probability, however, for the CBGF transmission targeting to the narrowband mMTC, the pilot collision is usually the dominant factor affecting the system performance. This is the design philosophy of ESOP.

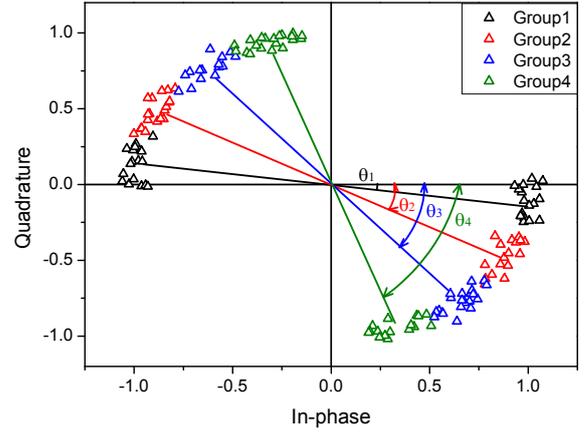

Fig. 4 Illustration of TO estimation with PM method, BPSK modulation

Generally speaking, the spatially combined data symbols $\hat{\mathbf{s}}_q$ are still distorted by the wireless channel and TO/FO. The estimation and compensation of these distortions are beyond the capability of the ESOP. Other efficient methods are required. Here, we introduce one simple method that utilizes the characteristics of the combined data symbols to estimate and compensate these distortions. For mMTC, the typical size of uplink payload is small and the IUI will be severe as the number of concurrent access UEs may be very large, therefore, low order modulation schemes like BPSK and QPSK are preferred. Utilizing the geometric characteristics of two-dimensional constellation, a partition-matching (PM) method proposed in our previous work [13] is used to estimate the wireless channel and TO/FO. Here, we take TO estimation as an example. Assume that the transmission bandwidth consists of $J$ subcarriers. TO will cause the constellation of the received modulation symbols to rotate in the frequency domain, as shown in Fig. 4. However, the regular geometric features of the modulation symbols are still reserved. To estimate the TO, $J$ subcarriers are divided into $G$ groups, with each group containing $J/G$ adjacent subcarriers. In each group, the average rotation angle θ of the received modulation symbols can be obtained using the PM method. Differentiating these average rotation angles, we get the differential angles, e.g. θ2-θ1, θ3-θ2, θ4-θ3, etc. Then, θ$_{avg}$ can be calculated by taking the average of these differential angles, and the rotation angle between two adjacent subcarriers caused by TO is θ$_{avg}$/(J/G). Finally, the TO compensation over the entire transmission resource can be performed. It should be noted that the estimated spatial channel at the position of non-zero element of the ESOP is accurate for UE without collision, thus this position requires no further compensation by the PM method and can be treated as the reference point for the TO compensation.

The method of FO estimation and compensation is similar, except that FO will cause the constellation of the received modulation symbols to rotate in the time domain, thus the OFDM symbols are divided into groups along the time domain direction. If TO and FO coexist, the estimation and compensation of TO and FO can be done alternately. Similar to TO estimation, the transmission bandwidth can also be divided into groups (or subchannels) to estimate the frequency selective fading channel. The difference is that the step of taking the average of these differential rotation angles in TO estimation is omitted, as the influence (e.g. rotation) of wireless channel on the constellation points is not as regular as the influence of TO.

TABLE I. SIMULATION PARAMETERS

| Parameter | Value |
|---|---|
| Carrier frequency | 700 MHz |
| System bandwidth | 10 MHz |
| Transmission bandwidth | 6 PRBs |
| Modulation and coding scheme | BPSK, LDPC |
| Size of transport block | 320 bits |
| Length of CRC | 16 bits |
| Pilot overhead | 1/7 |
| Code rate | 0.3889 |
| Antenna configuration | 1Tx, 8Rx |
| Channel model and delay spread | TDL-A 30ns, CDL-C 300ns |
| Time offset | Uniform within [0, 1CP] |
| Frequency offset | Uniform within $\pm 300$ Hz |

Then, the equalized data symbols are sent to the demodulator and decoder, and the cyclic redundancy check (CRC) is used to check whether the active UE is decoded correctly. The data symbols and pilots of correctly decoded UEs can be reconstructed and removed from the received data symbols $y_d$ and pilot symbols $y_P$. Next, MUD after successive IC is further performed until no active UE can be identified or no new UE can be successfully decoded. Since the data symbols of different UEs are uncorrelated, data-assisted channel estimation using the least-square method [13] can be used in the IC stage, which can improve the accuracy of channel estimation and reduce the residual error of IC.

## IV. NUMERICAL RESULTS

The performance of the proposed ESOP scheme is evaluated by link-level simulation, and comparisons with the TOP scheme are also provided. The simulation parameters are listed in Table I. The transmission resource is 6 consecutive PRBs in the frequency domain, and 14 OFDM symbols (1ms) in the time domain. The pilot overhead is 1/7, that is, 2 OFDM symbols are for pilot transmission. There are 12 subcarriers per PRB and the subcarrier spacing is 15 kHz. The size of data payload is 320 bits, and the code rate of LDPC is 0.3889. The BS is equipped with 8 receive antennas and each UE has one transmit antenna. Two channel models, tapped delay line model of type A (TDL-A) with delay spread of 30ns, and clustered delay line model of type C (CDL-C) with delay spread of 300ns, are adopted for evaluations and comparisons. For TDL-A 30ns channel model, the channel variation within the transmission bandwidth is small, and this channel model can be regarded as flat fading channel. On the contrary, CDL-C 300ns channel model can be regarded as frequency selective fading channel.

In the simulation, both single pilot scheme (TOP and ESOP scheme) and independent multi-pilot scheme (configured with 2 independent pilots) are considered. For TOP scheme and TOP scheme configured with 2 independent pilots, the size of pilot set is 24 and 12 respectively, and MMSE equalization using Eq. (5) is adopted. While for the proposed ESOP scheme and ESOP scheme configured with 2 independent pilots, 144 and 72 orthogonal pilots are available. To achieve better performance, the spatial combination vector in Eq. (5) is adopted for the proposed ESOP scheme, and Eq. (7) is used for the proposed ESOP scheme configured with 2 independent pilots.

### A. TDL-A 30ns Channel Model without TO/FO

Firstly, the BLER performance of TOP scheme and ESOP scheme is compared. As shown in Fig. 5, thanks to the greatly increased number of sparse orthogonal pilots and the decrease of pilot collision probability, the proposed ESOP scheme can support up to 30 simultaneous access UEs, which is twice that of TOP scheme.

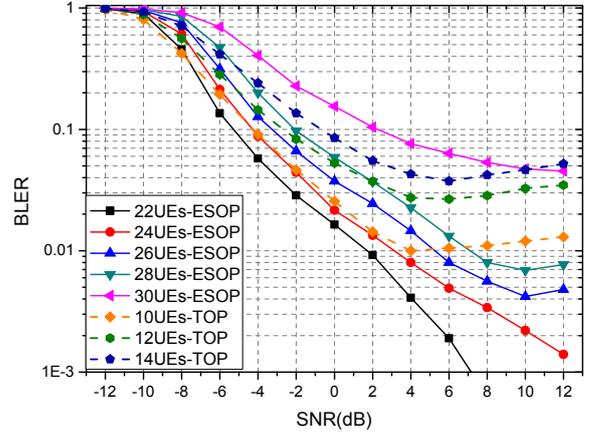

Fig. 5. The comparison of BLER performance between TOP scheme and the proposed ESOP scheme.

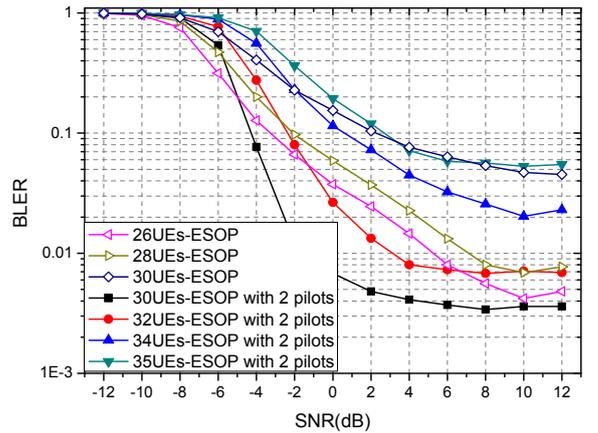

Fig. 6. The comparison of BLER performance of proposed ESOP scheme and ESOP scheme configured with 2 independent pilots.

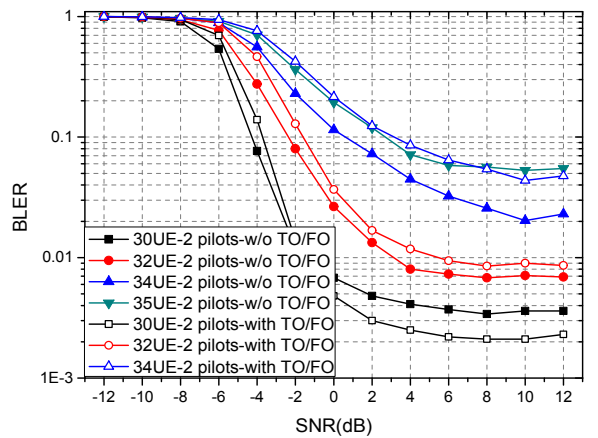

Fig. 7. The comparison of BLER performance of proposed ESOP scheme configured with 2 independent pilots under TDL-A channel with/without TO/FO.

Then, the BLER performance of ESOP scheme and ESOP scheme configured with 2 independent pilots is compared, as shown in Fig. 6. The ESOP scheme configured with 2 independent pilots can support up to 35 access UEs, which is 17% higher than the 30 UEs supported by ESOP scheme. At BLER = 0.1, the ESOP scheme configured with 2 independent pilots has a performance gain of 6 dB for 30 UEs, compared with the ESOP scheme. The better performance achieved by the ESOP scheme configured with 2 independent pilots corresponds to the lowest probability of pilot collision in Fig. 2.

*B. TDL-A 30ns Channel Model with TO/FO*

In this subsection, Both TO and FO are considered. Max TO is a normal CP length, which is about 4.69 μs [6]. In each transmission, the TO value of each UE's signal satisfies an independent uniform distribution in the interval [0, CP]. Similarly, FO value of each UE's signal satisfies an independent uniform distribution in the interval ± 300 Hz. The performance of ESOP configured with 2 independent pilots with/without TO/FO is compared. As shown in Fig. 7, the proposed ESOP scheme can still work well though large TO/FO exists. Specifically, the ESOP scheme configured with 2 independent pilots can support 34 access UEs with TO/FO, which is close to that without TO/FO. When the number of access UEs is less than 32, the performance loss caused by TO/FO is less than 1dB at BLER = 0.1.

*C. CDL-C 300ns Channel Model with TO*

In this subsection, the BLER performance of the TOP scheme and the proposed ESOP scheme is compared. Both schemes are configured with 2 independent pilots. The channel model is CDL-C 300ns, which can be considered as frequency selective fading channel, and TO within 1CP is considered. As shown in Fig. 8, thanks to the low pilot collision probability, the proposed ESOP scheme still performs better than TOP scheme under frequency selective fading channel. Specifically, the number of UEs supported by the proposed ESOP scheme doubles compared with TOP scheme.

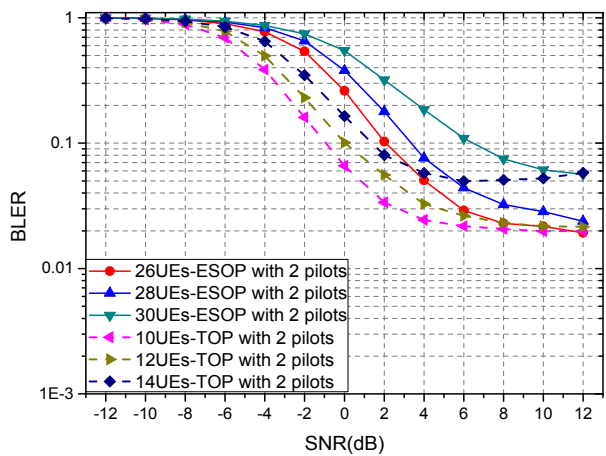

Fig. 8. The comparison of BLER performance between TOP and proposed ESOP scheme under CDL-C channel with 1CP TO, both schemes configured with 2 pilots. $J = 72$, $G = 24$, $J/G = 3$.

## V. CONCLUSION

Pilot collision has a dominant influence on the performance of grant-free transmission scheme. This paper proposed an extremely sparse orthogonal pilot scheme, which can greatly alleviate the pilot collision and exploit most part of the spatial capability. The simulation results show that the proposed ESOP scheme can significantly improve the BLER performance and increase the number of supported access UEs. The great performance improvement shows that the proposed ESOP scheme has potential applications in mMTC scenario in future wireless communications.